\documentstyle[aps,prl,floats,epsf,twocolumn]{revtex}

\preprint{IMSc-97/02/04}
\begin{document}
\twocolumn[\hsize\textwidth\columnwidth\hsize\csname
  @twocolumnfalse\endcsname
\draft 
\title{Time--of--night variation of solar neutrinos}
\author
{Mohan Narayan, G. Rajasekaran, Rahul Sinha} 
\address
{Institute of Mathematical Sciences, Madras 600 113, India.}
\date{21 February 1997}
\maketitle

\begin{abstract}

We investigate the time--of--night variation of solar neutrino rate
which will be of relevance to Super--Kamioka and Sudbury neutrino
detectors in the framework of oscillations among the three flavors. An
analytical method of computing the regeneration in the earth is presented. If
day-night effect is seen, we show how the study of the time--of--night
variation will allow the determination of the neutrino parameters.

\end{abstract}
\pacs{PACS number: 14.60.Pq, 95.30.Cq, 96.40.Tv, 96.60.Jw}
] 

It has been known for quite some time that an asymmetry between night
rate and day rate for the real time solar neutrino detectors is an
unambiguous signal for neutrino mixing and oscillations. Conversely
the absence of such an effect can put constraints on the neutrino
masses and mixing angles. Although no day--night asymmetry outside the
error--bars was seen at the Kamioka detector \cite{hirata,fukuda} the high
statistics detectors like Super-Kamioka \cite{totsuka},
SNO \cite{sno}and Borexino \cite{raghavan} will be much more effective
in investigating this effect. If there is a day-night asymmetry, then
the profile of the asymmetry as a function of the time during night is
a very sensitive function of the neutrino parameters. The counting
rates in these detectors are expected to be high enough for the study
of this time--of--night variation. The aim of this note is to focus
attention on this aspect of the day--night effect\cite{old}, in view
of the fact that Super--Kamioka has already started functioning and
SNO is expected to do so soon.

The neutrino samples different amounts of matter in the earth during a
single night and also during a year. The distance $d$ travelled by the
neutrino inside the earth during night,  as a function of time $t$,  is given by 
\begin{eqnarray}
&& d = 2 R (\sin \phi_{l} \sin \delta + \cos \phi_{l} \cos\delta \cos
(\frac{2\pi t}{T_D}))\, , \\
&&\sin \delta = \sin 23.5^o \sin (\frac{2\pi t}{T_{Y}}) ,
\end{eqnarray}
R is the radius of the earth, $\phi_{l}$ is the latitude of the
location of the detector, $T_D$ is the length of the day, $T_Y$ is the
length of the year and zero of $t$ is chosen at midnight on autumnal
equinox. Thus assuming that the neutrino parameters are in a suitable
range, neutrino data collected as a function of $t$ contain an
enormous amount of information on neutrino oscillations, which in
principle can be analyzed to yield the neutrino parameters.
\begin{figure}[htb]
\vskip -.5cm
\centerline{
\hbox  {
        \epsfxsize=4.5cm
        \epsfbox{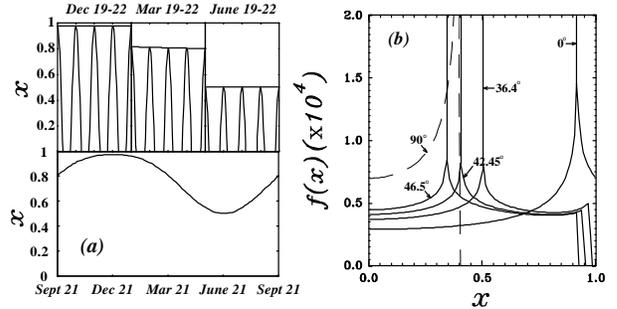}
       }
          }     
\caption{(a)The fractional distance $x$ travelled by the neutrino inside the earth during night
is plotted against time for the latitude $(36.4^o)$ of the Super--Kamioka detector.
The lower figure gives the envelope of the 365 maxima
during the year . As examples of the actual curves, those for a few nights during
three specific seasons of the year  are shown in the upper figure.
~(b)~
The function $f (x)$ in hours per unit $x$ is plotted for various latitudes: Super--Kamioka ($36.4^o$),
Borexino ($42.45^o$), SNO ($46.5^o$)  , equator ($0.0^o$) , pole ($90^o$).}
\label {Fig. 1}
\end{figure}

The time variation of $x=(\displaystyle\frac{d}{2 R})$ during the
night and year are illustrated in Fig.\ref{Fig. 1}a. If the data
collected during successive nights are accumulated at the
corresponding $x$-- bins, the calculation of neutrino rates per unit
bin will require the function $f (x)$ defined as the time duration per
unit interval of $x$ \cite{cherry}. $f(x)$ for different locations are
plotted in Fig.\ref{Fig. 1}b, which shows the relative merits of the
detectors for exposure to regions of $x$.


We now describe an analytical method of calculating the neutrino
regeneration effect in the  earth. Let a neutrino of  flavor $\alpha$
be produced at time $t=t_0$ in the core of the sun.
Its  state vector is 
\begin{equation}
| \Psi_{\alpha} (t_0) \rangle = |\nu_\alpha \rangle = \sum_i U^C_{\alpha i}
| \nu_i^C \rangle.
\end{equation}
where $|\nu_i^C \rangle$ are the  mass eigenstates
with mass eigenvalues $\mu_i^C$ and
$U^C_{\alpha i}$ are the elements of the  mixing
matrix in the core of the sun. We use Greek index $\alpha$ to denote the 
flavors e, $\mu$,$\tau$ and Latin index i to denote the mass eigenstates
i = 1,2,3. The neutrino propagates in the sun adiabatically upto $t_{R}$ (the
resonance point), makes nonadiabatic transitions at $t_R$, propagates
adiabatically upto $t_1$ (the edge of the sun) and propagates as a free particle
upto $t_2$ when it makes a nonadiabatic transition as it enters the earth. 
If it propagates adiabatically upto $t_{3}$
(we shall soon correct for nonadiabatic jumps during this propagation), the
probability amplitude for detecting a neutrino of flavor $\beta$ at $t_{3}$ is
\begin{eqnarray}
\langle \nu_\beta &&|\Psi_{\alpha} (t_3) \rangle = 
\sum_{k, j, i}
U^{E*}_{\beta k}  M^E_{k j}  M^S_{j i} U^C_{\alpha i}
exp \left\{-i\left(
\int_{t_2}^{t_3} \varepsilon_{k}^E (t) dt \right. \right. \nonumber \\
&& 
+\left. \left. \varepsilon_{j}
(t_{2}-t_{1}) + \int_{t_R}^{t_1} \varepsilon^{S}_{j} (t) dt + \int_{t_0}^{t_R}
\varepsilon^S_{i} (t)  dt 
\right)\right\}
\end{eqnarray}
$ \varepsilon_{i}  (\equiv E+\mu_i^{2}/2 E)$ ,$\varepsilon_{i}^{S} (t)$ and
$\varepsilon_{i}^{E} (t)$ are the energy eigenvalues in vacuum, sun and earth
respectively, 
$M^S_{j i}$ and $M^E_{j i}$ are  the probability amplitudes for
nonadiabatic transition $i\rightarrow j$ inside the sun and at the vacuum
-earth boundary respectively. The latter is due to the abrupt change in
density when the neutrino enters the earth and is given by
\begin{equation}
 M^E_{k j} = \langle \nu_{k}^E| \nu_{j} \rangle = \sum_{\sigma} \langle \nu_{k}^E|
\nu_{\sigma} \rangle \langle \nu_{\sigma}| \nu_{j}\rangle
    =   \sum_{\sigma} U^E_{\sigma k} U^*_{\sigma j}
\label{mekj}
\end{equation}
where $U$  and $U^E$ are  the mixing matrices  in vacuum and earth
respectively. 
Averaging the probability 
$|\langle \nu_\beta |\Psi_{\alpha} (t_3) \rangle|^2$ over $t_{R}$ results in the desired
incoherent mixture of mass eigenstates of neutrinos reaching the surface of the earth.
Calling this  
averaged probability as $P_{\alpha \beta}^N$ ( the probabilty for a
neutrino produced in the sun as $\nu_\alpha$ to be detected as
$\nu_\beta$ in the earth at night), we can write the result as
\begin{eqnarray}
 P_{\alpha \beta}^N &=& \sum_{j} P^S_{\alpha j} P^E_{j \beta} 
\label{PabN}\, , \\
 P^{S}_{\alpha j} & = & \sum_i |M^S_{j i}|^2 |U^C_{\alpha
i}|^2 \, , \\ \label{PEjb}
P^E_{j \beta} & = & 
\sum_{k, k'} U^{E*}_{\beta k} U^{E}_{\beta k'} M^E_{k j} 
M^{E*}_{k' j} exp \left(-2 i \Phi_{k k'}\right)\, , \\
\Phi_{k k'} &=&\frac {1}{2} \int_{t_2}^{t_3}
( \varepsilon_{k}^E (t) - \varepsilon_{k'}^{E} (t)) dt \, .
\end{eqnarray}
For the daytime, put $t_{3} = t_{2}$ so that
$P^E_{j \beta}$ becomes $|U_{\beta j}|^2$ and so eq.(\ref{PabN})
reduces to the usual \cite{parke,narayan} transition probability in the day:
\begin{equation}
P^{D}_{\alpha \beta} = \sum_i \sum_j |U_{\beta j}|^2 |M^S_{j i}|^2
|U^{S}_{\alpha i}|^2 \, .
\label{pdab}
\end{equation}

It is important to note that the factorization of probabilities seen in
eq.(\ref{PabN}) is valid only for mass--eigenstates in the
intermediate state. An equivalent statement of this result is that the
density matrix is diagonal only in the mass-eigenstate representation
and not in the flavor representation. 

We next show how to take into account nonadiabatic jumps during the
propagation inside the earth. Consider $\nu$ propagation through a
series of slabs of matter, density varying inside each slab smoothly
but changing abruptly at the junction between adjacent slabs. The
state vector of the neutrino at the end of the $n^{th}$ slab $|n
\rangle$ is related to that at the end of the $(n-1)^{th}$ slab $|n-1
\rangle$ by $|n \rangle = F^{(n)} M^{(n)} |n-1 \rangle$ where $M
^{(n)}$ describes the nonadiabatic jump occuring at the junction
between the $(n-1)^{th}$ and $n^{th}$ slabs while $F^{(n)}$ describes
the adiabatic propagation in the $n^{th}$ slab. They are given by
\begin{eqnarray}
M^{(n)}_{i j} &=& \langle \nu_{i}^{(n)} | \nu_{j}^{(n-1)} \rangle  = (U^{(n)^ {\dagger}}
U^{(n-1)})_{i j}^*  \, , \label{Mn}\\
F^{(n)}_{i j} &=& \delta_{i j} exp \left(-i \int_{t_{n-1}}^{t_{n}} \varepsilon_{i}
(t) dt \right)\, ,
\end{eqnarray}
where the indices $(n)$ and $(n-1)$ occuring on $\nu$ and $U$  refer 
respectively to the $n^{th}$ and $(n-1)^{th}$ slabs at the junction between these 
slabs. Also note that $M^{(1)}$ is the same as $M^{E}$ defined in eq(\ref{mekj}). Defining
the density matrix at the end of the $n^{th}$ slab as
$\rho^{(n)} = | n \rangle \langle n |$ , we have the recursion formula
\begin{equation}
\rho^{(n)} = F^{(n)} M^{(n)} \rho^{(n-1)} M^{(n)\dagger}
F^{(n)\dagger}.
\label{rho}
\end{equation} 
Starting with
$\rho^{(0)} = | \nu_{j} \rangle \langle \nu_{j} |$ (i.e $\nu_j$  entering the earth),
 we can calculate $\rho^{(N)}$ at the end of the $N^{th}$ slab using (\ref{rho}).
The probability of observing $\nu_\beta$ at the end of the $N^{th}$ slab is
\begin{equation}
P^{E}_{j \beta} = \langle \nu_{\beta} | \rho^{(N)} | \nu_{\beta} \rangle = ( U^{(N)}
\rho ^{(N)^*}  U^{(N)^ {\dagger}} )_{\beta \beta} \, .
\end{equation}
This formula (which reduces to eq(\ref{PEjb}) for $N$ = 1) 
can be used for the earth
modelled as  consisting of $(N+1)/2$ concentric shells, with the density
varying gradually within each shell. We shall present numerical results
for $N = 3$ (mantle and core) \cite{stacey}. However for $x < 0.84$, neutrinos pass
only through the mantle and so $N = 1$. Accuracy achieved with this model
is adequate for the present purposes, but the formalism allows one to improve
the accuracy to any desired level, by adding more shells.

Apart from the nonadiabatic jumps occuring at the
density--discontinuities, such jumps can occur also at any MSW
resonance in the earth. The formalism presented above is capable of
handling this. One simply replaces eq.(\ref{Mn}) for $M^{(n)}$ for that
transition by an appropriate Landau--Zener formula\cite{kuopan}.

\begin{figure*}[htb]
\centerline{
\hbox  {
        \epsfxsize=10.cm
        \epsfbox{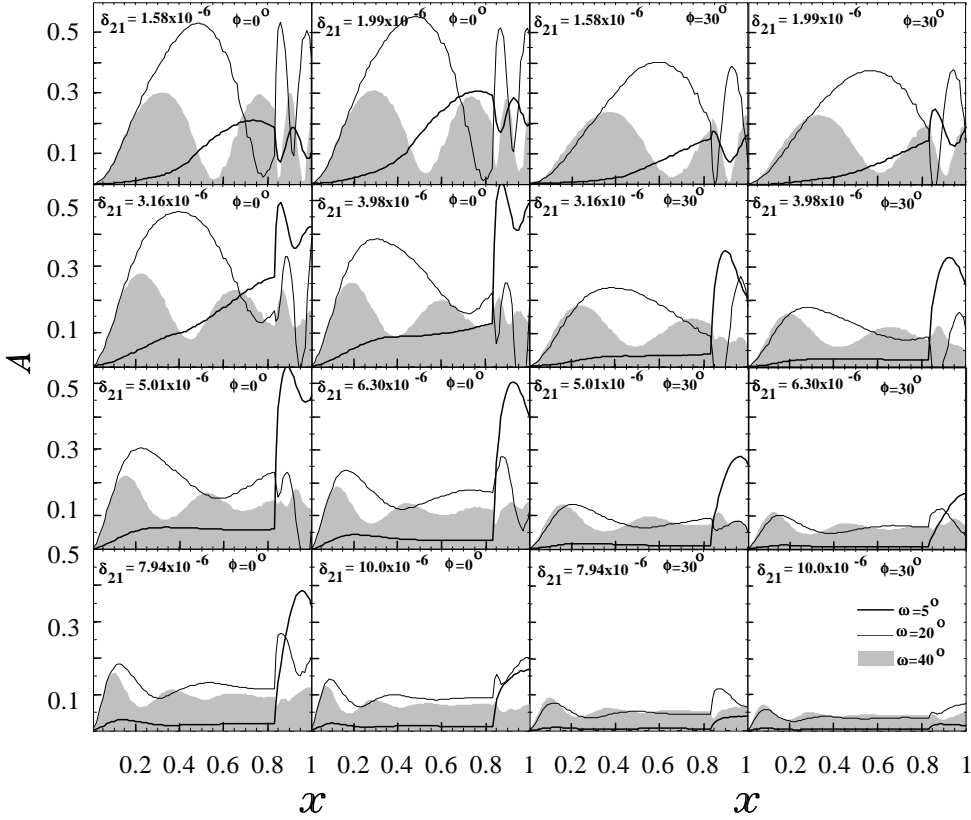}
       }
          }     
\caption{
$A$ as a function of $x$,  the fractional distance travelled by the 
neutrino, for stated values of $\omega$, $\delta_{21}$ (in $eV^2$)
and $\phi$.
}\label {Fig. 2}
\end{figure*}

We parametrize the mixing matrix U in vacuum as
$U = U^{23} (\psi) U^{13} (\phi) U^{12} (\omega)$
where $U^{ij} (\theta_{ij})$ is the two flavor mixing matrix between the ith and the
jth mass eigenstates with the mixing angle $\theta_{ij}$, neglecting CP violation.
In the solar neutrino problem $\psi$ drops out \cite{kuopan,ajmm}.   
The mass differences in
vacuum are defined as $\delta_{21} = \mu^{2}_{2} - \mu_{1}^{2}$ and
$\delta_{31} = \mu^2_{3} - \mu^2_{1}$. It has been shown \cite{narayan,fogli} 
that the
simultaneous solution of both the solar and the atmospheric neutrino
problems requires the mass hierarchy $\delta_{31} \gg \delta_{21}$
 and under this condition $\delta_{31}$ also drops out.
The rediagonalization of the mass matrix in the presence of matter (in the
sun   or earth) under the hierarchy condition  leads to the following results \cite{narayan}
\begin{eqnarray}
&&\tan 2 \omega_m  =  \frac{\delta_{21} \sin 2 \omega}{
		  \delta_{21} \cos 2 \omega - A \cos^2 \phi}
		  \label{eq13}\, , \label{eq15}\\ 
&&\sin \phi_{m} = \sin \phi  \label{eq14}\, \label{eq16}, \\
&&\delta_{21}^{m} = \delta_{21} \cos 2 (\omega - \omega_{m})- A \cos^2\phi \cos 2 
\omega_{m} \, ,
\label{eq17}
\end{eqnarray} 
where A is the Wolfenstein term $A = 2 \sqrt{2}~G_F~N_e\,E$ ($N_{e}$
is the number density of electrons and E is the neutrino energy) .
 We note that $\delta_{31}\gg A$, for $A$ evaluated at any
point in the sun or the earth. In eqs (\ref{eq15}) - (\ref{eq17}), the ``$m$'' stands for
matter and in using these equations, one must use the appropriate density of
matter that is required at the various points along the trajectory of the neutrino.

All the probabilities $P^{N}_{\alpha \beta}$, $P^S_{\alpha j}$ and $P^E_{j \beta}$
satisfy the normalization conditions, as for instance, $\sum_{j} P^E_{j \beta} = 1$.
For three flavors, use of these conditions allows us to express $P^N_{e e}$ in terms
of $P^D_{e e}$, $P^E_{1 e}$, $P^E_{2 e}$ and $P^S_{e 3}$ as
\begin{eqnarray}
&&P^{N}_{e e} = [P^D_{e e} (P^E_{1 e} - P^E_{2 e}) - \cos^{2}\phi (\sin^2\omega   P^{E}_
{1 e} - \cos^2\omega   P^{E}_{2 e})   \nonumber\\ 
&& - P^{S}_{e 3}((3 \cos^2\phi \cos^2\omega -1) P^{E}_{2 e} - (3 \cos^2\phi \sin^2\omega
- 1) P^{E}_{1 e}\nonumber \\
&& - \cos^2\phi \cos^2\omega)]/(\cos^2\phi \cos2 \omega).
\label{PNee}
\end{eqnarray}
This is a general formula that goes over to the one given in
Ref. \cite{mksm} for two flavors. Simplifications arise from the mass heirarchy
condition. 
First, $M^{S}_{i j}$ is nonzero for $i,j = 1,2$ only and hence we can replace 
$P^{S}_{e 3}$ in eq(\ref{PNee}) by $\sin^2\phi_{C}$. $|M^{S}_{1 2}|^2$ is taken to be
the modified Landau- Zener jump probability for an exponentially varying
solar density $\cite{kuopan}$. Further, $M^{(n)}$ also are reduced to $2 \times 2$ matrices. 
In fact eq(\ref{Mn}) gives $M^{(n)} = U^{1 2} (\theta)$ where $U^{1 2} (\theta)$ is the
2--flavor mixing matrix, with mixing angle $\theta = \omega_{n} - \omega_{n-1}$.
As a result, we get the simple formulae from eq(\ref{PEjb}), valid for $x < 0.84$ :
\begin{eqnarray}
P^{E}_{ 1 e} &=& \cos^2\phi [\cos^2\omega_{E} - \sin 2 \omega_{E} \sin 2 (\omega_{E} -
\omega) \sin^{2} \Phi_{1 2} ]          \\ 
P^{E}_{ 2 e} &=& \cos^2\phi [\sin^2\omega_{E} + \sin 2 \omega_{E} \sin 2 (\omega_{E} -
\omega) \sin^{2} \Phi_{1 2} ]           
\end{eqnarray} 
where $\omega_{E}$ is the mixing angle, just below the surface of the earth.

The neutrino detection rates for a Super--Kamioka type of  detector is
given by
\begin{equation}
R =\int\phi\sigma  P_{ee} dE +\frac{1}{6}
 \int \phi \sigma  (1-P_{ee}) dE  
\end{equation}
where the second term 
is the  neutral current contribution and $\phi(E)$ is the solar
neutrino flux as a function of the neutrino energy $E$ and
$\sigma(E)$ is the cross section from neutrino electron scattering and
we integrate from $5MeV$ onwards. The
cross section is taken from  \cite{bahcall} and the flux from
\cite{bahpin92}. The rates for the night and day $R_N$ and
$R_D$ are calculated using $P_{ee}^N$ and $P_{ee}^D$ respectively. 
We define the day--night asymmetry ratio as  $A=(R_N-R_D)/(R_N+R_D)$.
We can multiply $R_{N}$ and $R_{D}$ by the function $f (x)$ displayed
in Fig.\ref{Fig. 1} to get the rates per unit interval in $x$. Note that
$f(x)$ cancels in the asymmetry ratio calculated
theoretically. However, the experimentalists have to weight the day
rate $R_{D}$ with $f (x)$ before comparing their data with our
theoretical curves.

\begin{figure}[htb]
\centerline{
\hbox  {
        \epsfxsize=4.5cm
        \epsfbox{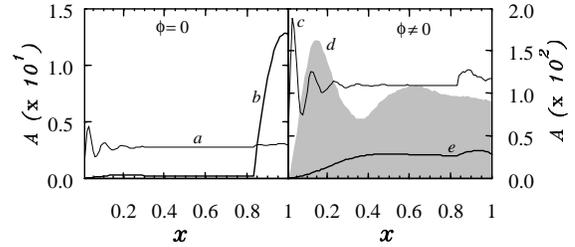}
       }
          }     
\caption{ $A$ as a function of $x$, for (a) $\delta_{21}$ = $3 \times
10^{-5}$ $eV^2$, $\omega$ = $28^o$, $\phi$ = 0; (b) $\delta_{21}$ =
$6\times 10^{-6}$ , $\omega$ = $2.5^o$, $\phi$ = 0; (c)
$\delta_{21}$ =  $2.5\times 10^{-5}$ , $\omega$ = $20^o$, $\phi$
= $32^o$; (d) $\delta_{21}$ = $6.3\times 10^{-5}$ , $\omega$ =
$18^o$, $\phi$ = $45^o$; (e) $\delta_{21}$ =  $2.5\times 10^{-6}$,
$\omega$ = $6^o$, $\phi$ = $50^o$.  }\label {Fig. 3}
\end{figure}

In Fig.\ref{Fig. 2} and \ref{Fig. 3} we have plotted $A$ as a function of $x$, the  
fractional distance travelled by the neutrino inside the earth
 for various values of the neutrino parameters
$\delta_{21}, \omega$ and $\phi$. 
Different values of these parameters
have distinguishable characteristics. Some gross features which may enable
us to specify their approximate domains 
 are the following:
\begin{itemize}
\item For small angle $\omega$ there is a gradual increase of the
  asymmetry with $x$, whereas for large $\omega$ the oscillations in
  $x$ start showing up. For $x<0.84$ ({\it i.e.} trajectories through
  mantle only) there is a very clear discrimination between the small
  $\omega$ and large $\omega$, irrespective of $\delta_{21}$ and
  $\phi$.
\item As $\phi$ increases, the asymmetry at any $x$ decreases.
\item The amplitude
of the oscillatory pattern is largest for small $\delta_{21}$ and
decreases steadily as $\delta_{21}$ increases.
\item For small $\omega$ and large $\delta_{21}$, asymmetry is
appreciable only in the core and is a sensitive function of $\delta_{21}$.
\end{itemize}

For non--zero $\phi$ a fraction of the solar neutrinos come out of the
sun as $\nu_{\tau}$ and these $\nu_{\tau}$
cannot reconvert back to $\nu_{e}$ inside earth because $\phi$ is
not affected by matter (see eqn.(\ref{eq14})). So a non zero $\phi$
dilutes the asymmetry. Our numerical results include the effect of any
adiabatic MSW resonances that may occur inside the earth. For
$\phi=0$, as pointed out recently \cite{blwz,Rosen}, MSW resonances do occur
in the earth's core. However, for large $\phi$ they disappear and this
is another reason for the regeneration efect to be smaller for large
$\phi$, in the core. 

In Fig. \ref{Fig. 3} we have chosen a few parameter sets for which $A$
is very small $(<0.15)$ since they are possible solutions to the solar
and atmospheric neutrino problems \cite{narayan,fogli,Langacker}. But
we cannot rigorously exclude other values of the neutrino parameters
at the present stage of knowledge. Day--night effect must be
studied in an unbiased manner, especially because the ratio A is
relatively independent of the uncertainties of the solar models.



After this work was completed, we came to know of a related work:
E. Lisi and D. Montanino, hep-ph/9702343. They have also stressed the
importance of an analytic approach, however their method is different
from ours and they have ignored the third flavor.

We thank M. C. Sinha, M. V. N. Murthy and S. Uma Sankar for
discussions.

\end{document}